\documentclass[11pt]{article}
\usepackage{osid}
\usepackage{graphicx}
\usepackage{colordvi}

\title{Effective generation of cat and kitten states}
\author{Magdalena Stobi\'nska \\{\footnotesize\it Instytut Fizyki
    Teoretycznej, Uniwersytet Warszawski, Warszawa 00--681, Poland \&
    magda.stobinska@fuw.edu.pl}\\[2ex] G. J. Milburn
  \\{\footnotesize\it Centre for Quantum Computer Technology and
    School of Physical Sciences, The University of Queensland, St
    Lucia, Queensland 4072, Australia \&
    milburn@physics.uq.edu.au}\\[2ex] Krzysztof W\'odkiewicz
  \\{\footnotesize\it Instytut Fizyki Teoretycznej, Uniwersytet
    Warszawski, Warszawa 00--681, Poland \\ and Department of Physics
    and Astronomy, University of New Mexico, Albuquerque,
    NM~87131-1156, USA \& wodkiew@fuw.edu.pl} }

\begin{document}

\maketitle
\begin{abstract}
     We present an effective method of coherent state superposition
     (cat state) generation using single trapped ion in a Paul
     trap. The method is experimentally feasible for coherent states
     with amplitude $\alpha \le 2$ using available technology.  It
     works both in and beyond the Lamb-Dicke regime.
\end{abstract}

\section{Introduction}

One of the most inspiring aspects of quantum physics is the
possibility of generation a quantum superpositions of two more
macroscopic, classically distinguishable, interfering states.  This
idea is closely related to the famous Schr\"odinger cat paradox
\cite{Schroedinger}, where the cat is set to be alive and dead with
equal probabilities until the measurement is made. This state is
entangled to the device that can kill the cat. In recent literature
just the superposition of two coherent states with a $\pi$ phase
difference and a large amplitude inherited the name, and is referred
as a cat state. A superposition of more than two coherent states is
called a kitten state.

The cat and kitten states have brought a lot of interest of physicists
due to many possible applications in quantum information processing
\cite{vanEnk, Jeong, Jeong2, Ralph2003, Ralph2}, quantum teleportation
\cite{vanEnk, Jeong}, quantum nonlocality tests \cite{Jeong3, Buzek},
generation and purification of entangled coherent states \cite{Jeong},
and quantum computation and communication \cite{Jeong, Jeong2,
  Ralph2003, Glancy}.  Those states are also very useful for
investigation of the decoherence process \cite{Monroe, Myatt2000}.

So far, a superposition of two coherent states has been successfully
generated for phonon modes of a single trapped ion \cite{Monroe} and
superconducting cavity \cite{Brune}.  A lot of effort has been made to
investigate the possibility of photon coherent state superposition
generation using Kerr nonlinearity \cite{Jeong4, Stobinska}.  However,
the nonlinearity is far too small to ensure the effective method of
the state preparation.

The aim of this article is to focus on the possibilities of the
coherent state superposition generation, which are offered by the
trapped ions in a Paul trap. The ion traps have generated a lot of
interest due to their possible applications in quantum information
theory~\cite{Blatt2005} and quantum computation~\cite{Ekert1996}.  In
different experiments, Fock number states~\cite{Roos1999}, coherent
states~\cite{Meekhof1996}, vacuum squeezed states~\cite{Heinzen1990},
and Schr\"odinger cat states~\cite{Monroe} has been
realized. According to our knowledge, neither non-Gaussian states
(other than cat state) nor a superposition of more than two coherent
states have been observed so far.

The presented method allows for the generation of two and more,
eg. six, coherent states superposition. It is closely related to the
Kerr nonlinear interaction. It is experimentally feasible for coherent
states with the amplitude $\alpha \le 2$.

\section{Kerr state}

Let us begin the discussion with a brief summary of a Kerr state and
a Kerr medium.

The one-mode Kerr state results from an interaction of a coherent
state of light $|\alpha\rangle$ with a third-order $\chi^{(3)}$
nonlinear medium, the Kerr medium \cite{Tanas}. The optical fibers are
the best known example of the Kerr medium.

The Hamiltonian describing the interaction in the ideal medium,
without damping and thermal noise, is of the following form
\begin{equation}
H = \hbar \frac{\kappa}{2}\, a^{\dagger}a^{\dagger}a\, a,
\end{equation}
where $a$ and $a^{\dagger}$ are annihilation and creation operators of
the light mode. The strength of the medium is given by the nonlinear
constant $\kappa = \frac{8 \pi^2 \hbar \omega^2}{\epsilon_0
  n_0^4(\omega)V} \chi^{(3)}$, where $\omega$ is the frequency of the
injected light beam, $n_0(\omega)$ is the linear refractive index and
$V$ is the volume of quantization.

 The one-mode Kerr state is an infinite superposition of different
 photon number states (Fock states)
\begin{equation}
|\Psi_K(\tau)\rangle = e^{-\frac{i H t}{\hbar}}|\alpha\rangle
= e^{-\frac{|\alpha|^2}{2}} \sum_{n=0}^{\infty}\,
 \frac{\alpha^n}{\sqrt{n!}}\, e^{i\frac{\tau}{2}n(n-1)}
 |n\rangle.
\label{Kerr}
\end{equation}
Its properties are characterized by a dimensionless parameter $\tau=
-\kappa t$, where $t$ is a time that the light has spent in the fiber.

\begin{figure}[h!]
\begin{center}
\scalebox{0.7}{\includegraphics{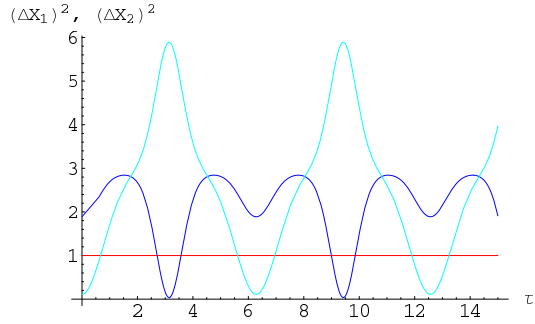}}
\caption{(Color online) The evolution of the amplitude and the phase
  quadrature uncertainties, $(\Delta X_1)^2$ (dark blue) and $(\Delta
  X_2)^2$ (light blue), for $\alpha=1$. The line is a reference value
  of $1$ obtained for a vacuum.}
\label{quadratures}
\end{center}
\end{figure}

Although the statistics of the Kerr state is Poissonian,
$g^{(2)}(\tau)=1$, this is a squeezed state. This fact can be easily
observed from the evolution of its electric field quadrature
uncertainties
\begin{eqnarray}
(\Delta X_1)^2 &=& 1+ 2 \alpha^2 \left( 1- e^{2\alpha^2(\cos \tau - 1)}
+ \mathrm{Re} \{e^{i\tau + \alpha^2(e^{2i\tau - 1})} - e^{2\alpha^2(e^{i\tau-1})}\}
\right),
\\
(\Delta X_2)^2 &=&  1+ 2 \alpha^2 \left( 1- e^{2\alpha^2(\cos \tau - 1)}
- \mathrm{Re} \{e^{i\tau + \alpha^2(e^{2i\tau - 1})} - e^{2\alpha^2(e^{i\tau-1})}\}
\right),
\end{eqnarray}
where $X_1 = a + a^{\dagger}$ and $X_2 = -i(a - a^{\dagger})$ are
amplitude and phase quadratures.  Depending on $\tau$, the quadratures
are squeezed alternately: the quantum fluctuations in one quadrature
are reduced below the vacuum level, $1$, at the expense of increased
fluctuations in the other one, see Fig.\ref{quadratures}. If none of
them is squeezed for a given value of $\tau$, the squeezing will not
be seen in the principal axes directions but for quadratures at a
certain angle. This fact corresponds to the rotation of the error
contour in the phase space.

The Kerr state is a non-Gaussian quantum state. This information can
be obtained from its Wigner phase space function: it takes the
negative values and it is not rotational symmetric. The Wigner
function can be computed and expressed in two equivalent ways
\begin{eqnarray}
W(\tau, \gamma, \gamma^*) &=& \frac{2}{\pi}\, e^{-2|\gamma|^2}
e^{-|\alpha|^2}\, \sum_{q=0}^{\infty} \frac{\left(2\alpha^{*}\gamma
  e^{i\frac{\tau}{2}}\right)^q}{q!}  e^{-i\frac{\tau}{2}q^2}
\nonumber\\ &\times&
\sum_{k=0}^{\infty}\frac{\left(2\alpha\gamma^*e^{-i\frac{\tau}{2}}\right)^k}
    {k!} e^{i\frac{\tau}{2}k^2} e^{-|\alpha|^2
      e^{i\tau(k-q)}},\label{summation}\\ W(\tau, \gamma, \gamma^*)
    &=& \frac{2}{\pi}\, e^{2|\gamma|^2}
    e^{-|\alpha|^2}\,\sum_{n,m=0}^{\infty} \frac{1}{(-2)^{n+m}}
    \frac{\alpha^n}{n!}\frac{{\alpha^*}^m}{m!} \nonumber\\ &\times&
    e^{i\frac{\tau}{2}\left[n(n-1)-m(m-1)\right]}
    \left(\partial_{\gamma}\right)^n
    \left(\partial_{\gamma^*}\right)^m
    e^{-4|\gamma|^2}.\nonumber\\ {}\label{derivatives}
\end{eqnarray}

\begin{figure}[h!]
\scalebox{0.7}{\includegraphics{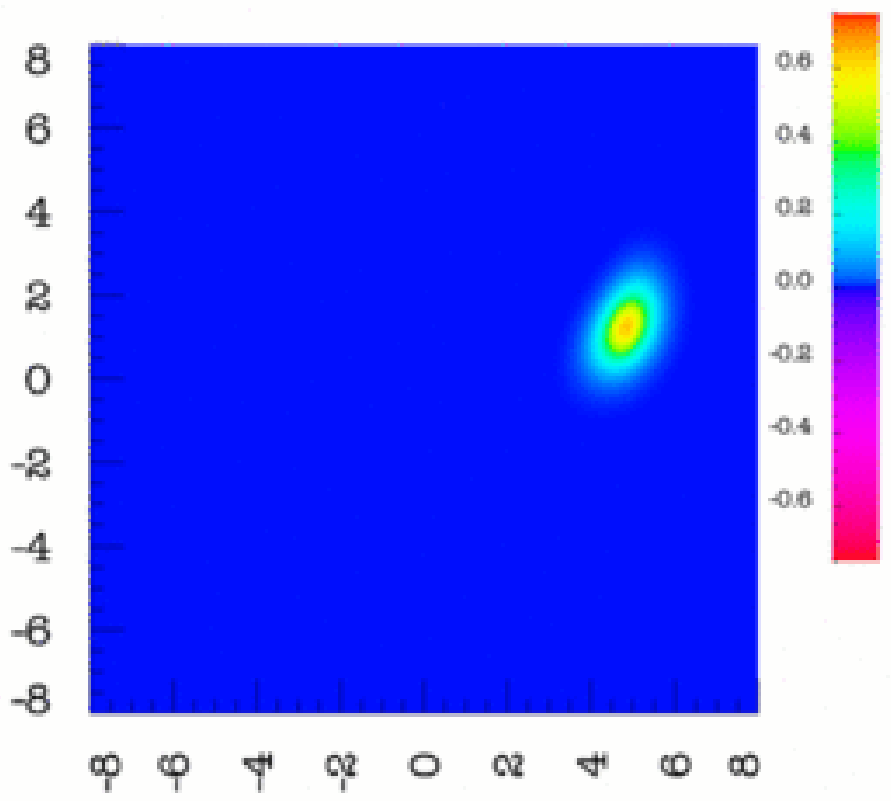}}
\scalebox{0.7}{\includegraphics{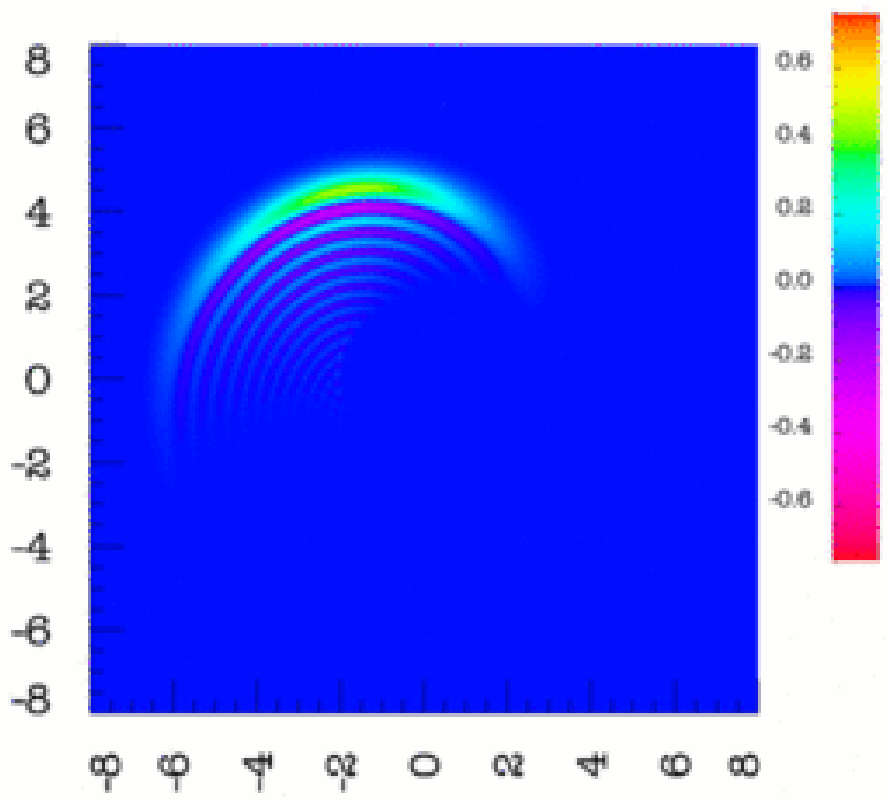}}
\caption{(Color online) The Wigner function evaluated for $\alpha=5$.
  For $\tau=0.01$ the Wigner function is an ellipse and the state
  becomes squeezed - the left figure. The Wigner function for $\tau =
  0.08$ - the right figure.  The negativities appear and form a tail
  of interference fringes.}
\label{Wigner_a_5}
\end{figure}

The examples of the Wigner function evaluated for $\alpha=5$ and two
values of evolution parameter $\tau = 0.01$ and $\tau=0.08$ are
depicted on Fig. \ref{Wigner_a_5}. The distribution starting with a
circular shape genuine to a coherent state turns into an ellipse and
the state becomes squeezed - the left figure. It also shows that the
Kerr state approximates a one-mode Gaussian squeezed state for $\tau
\simeq 0$ very well. Then, the ellipse is stretched into a banana
shape - the right figure. This plot reveals the nonclassicality of the
state: the negative values of the Wigner function form a tail of
interference fringes following the ``main'' part of the ``banana''
distribution.

\section{Cat and kitten states}

The evolution of a coherent state in a Kerr medium is periodic: the
phase factor in the Kerr state (\ref{Kerr}) is a periodic function of
the parameter $\tau$. Therefor, we achieve the same state for $\tau$
and $\tau + 2\pi$. Moreover, if $\tau$ is taken as a fraction of the
period of the evolution, $\tau = 2\pi R$ where $R<1$ is a rational
number, the infinite sum of Fock states breaks into a finite sum of
coherent states, all of the same amplitude but different phases
\cite{Bialynicka}. This effect is also known as a fractional
revival~\cite{Averbukh1989}.

\begin{figure}[h!]
\begin{center}
\hbox to \textwidth{
\hss
\scalebox{0.7}{\includegraphics{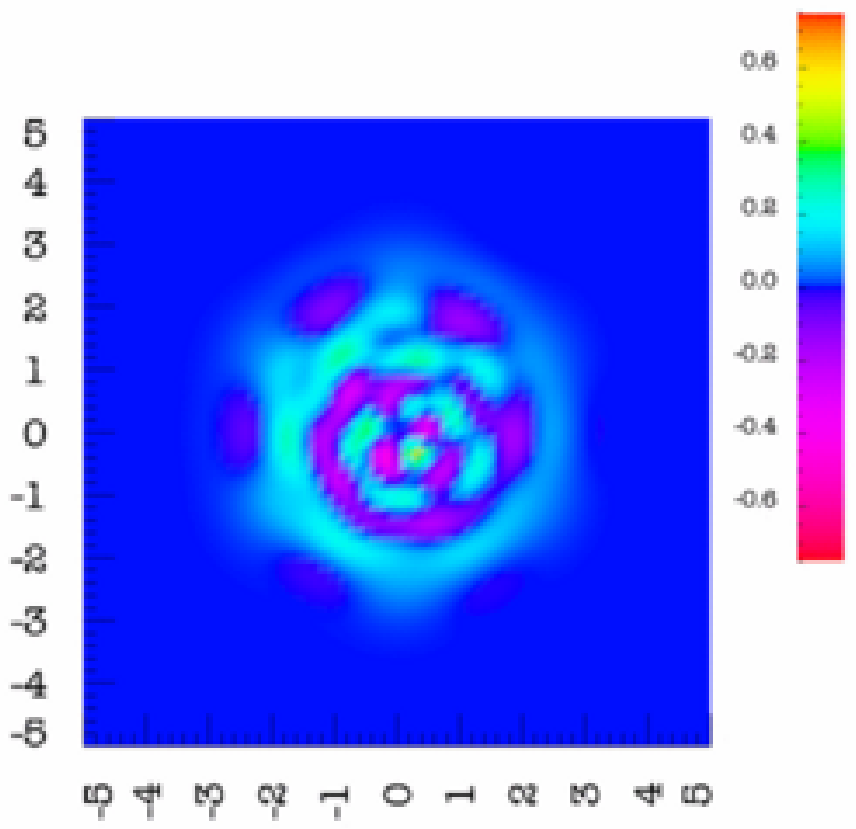}}
\scalebox{0.7}{\includegraphics{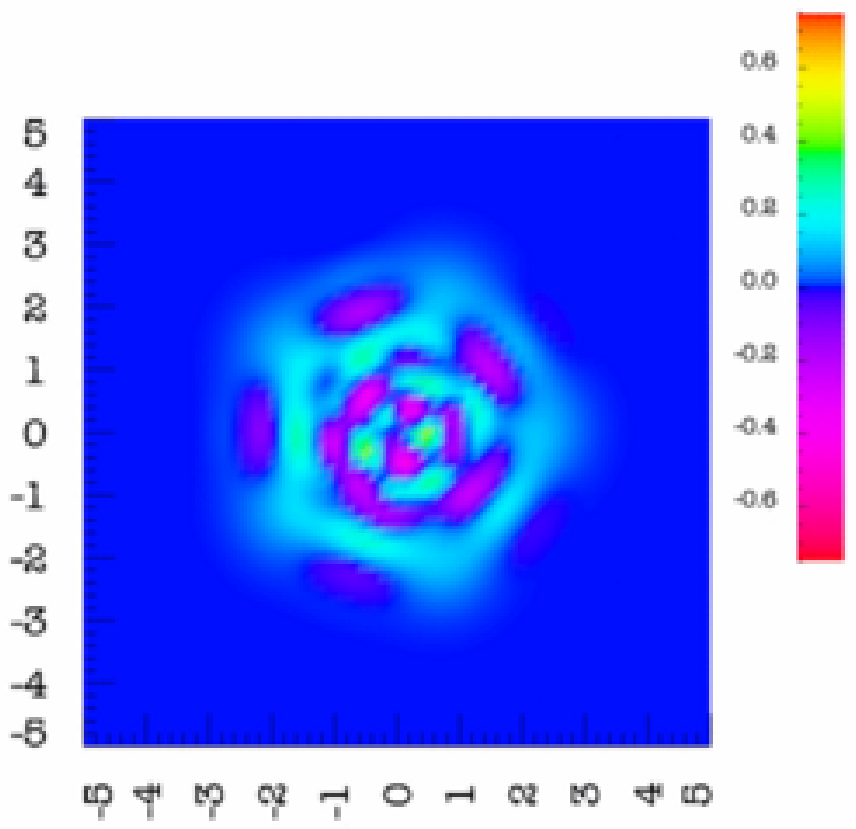}}
\hss}
\caption{(Color online) The Wigner function evaluated for $\alpha=
  2$. The left figure: $\tau=\frac{\pi}{3}$. The Kerr state becomes a
  superposition of six coherent states: $|2e^{i\frac{\pi}{6}}\rangle$,
  $|2e^{i\frac{\pi}{2}}\rangle$, $|2e^{i\frac{5\pi}{6}}\rangle$,
  $|2e^{i\frac{7\pi}{6}}\rangle$, $|2e^{i\frac{3\pi}{2}}\rangle$,
  $|2e^{i\frac{11\pi}{6}}\rangle$. The right figure:
  $\tau=\frac{2\pi}{5}$.  The Kerr state becomes a superposition of
  five coherent states: $|2\rangle$, $|2e^{i\frac{2\pi}{5}}\rangle$,
  $|2e^{i\frac{4\pi}{5}}\rangle$, $|2e^{i\frac{6\pi}{5}}\rangle$,
  $|2e^{i\frac{8\pi}{5}}\rangle$.}
\label{Wigner_1}
\end{center}
\end{figure}

\begin{figure}[h!]
\begin{center}
\hbox to \textwidth{
\hss
\scalebox{0.7}{\includegraphics{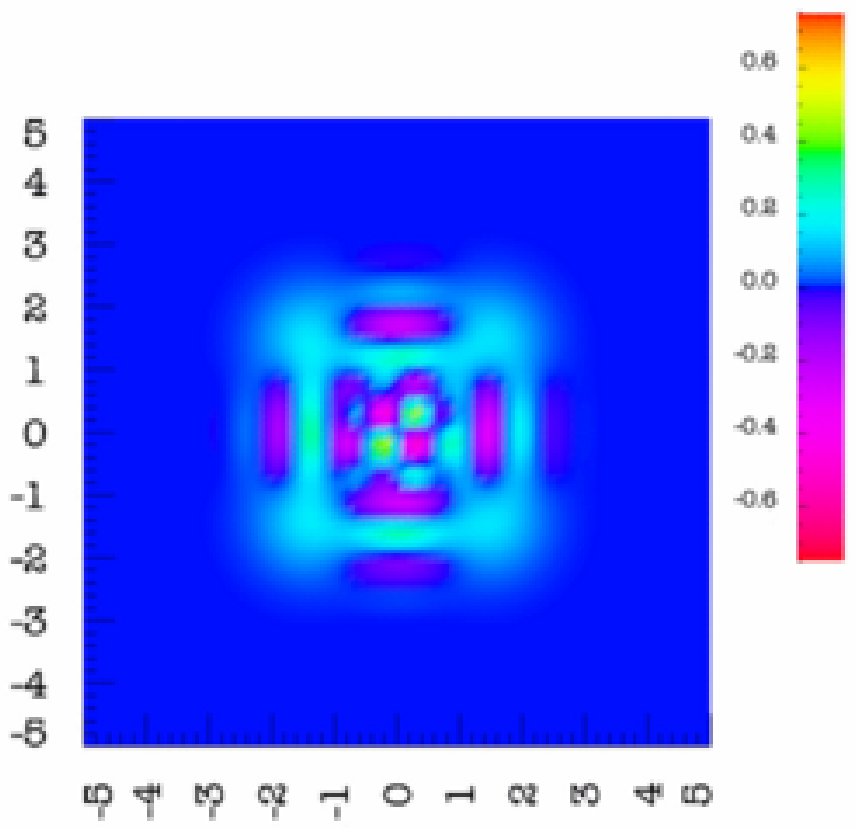}}
\scalebox{0.7}{\includegraphics{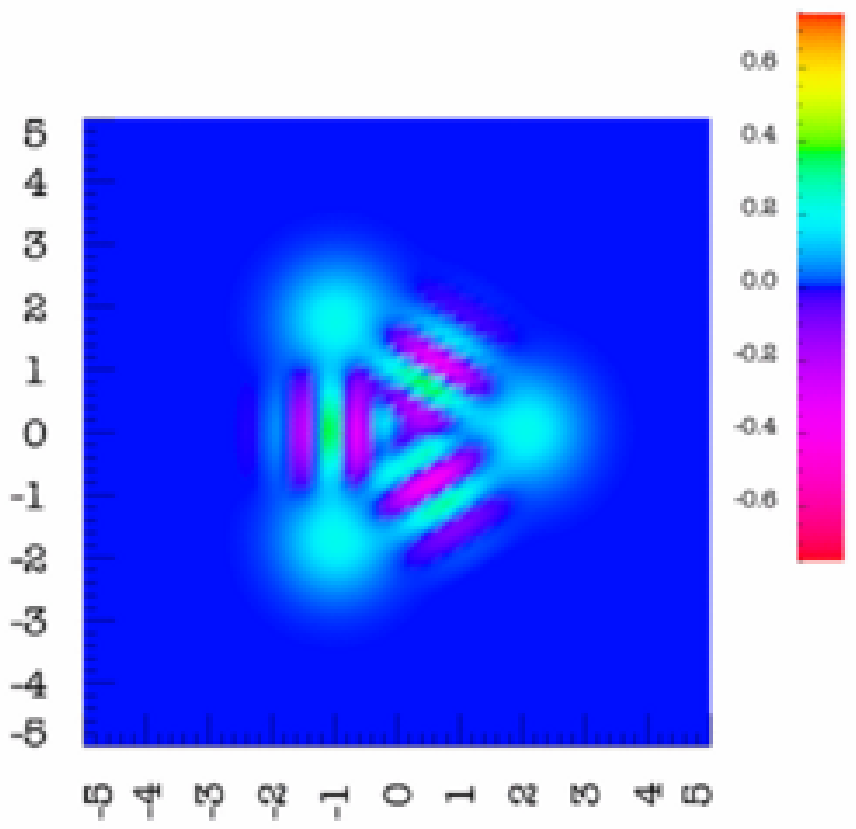}}
\hss}
\caption{(Color online) The Wigner function evaluated for $\alpha=
  2$. The left figure: $\tau=\frac{\pi}{2}$. The Kerr state becomes a
  superposition of four coherent states:
  $|2e^{i\frac{\pi}{4}}\rangle$, $|2e^{i\frac{3\pi}{4}}\rangle$,
  $|2e^{i\frac{5\pi}{4}}\rangle$, $|2e^{i\frac{7\pi}{4}}\rangle$. The
  right figure: $\tau=\frac{2\pi}{3}$. The Kerr state becomes a
  superposition of three coherent states: $|2\rangle$,
  $|2e^{i\frac{2\pi}{3}}\rangle$, $|2e^{i\frac{4\pi}{3}}\rangle$.}
\label{Wigner_2}
\end{center}
\end{figure}

\begin{figure}[h!]
\begin{center}
\scalebox{0.7}{\includegraphics{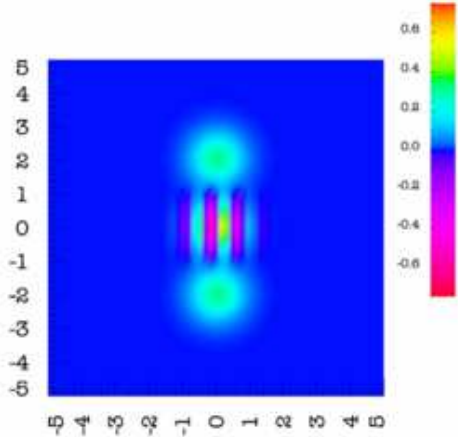}}
\caption{(Color online) The Wigner function evaluated for $\alpha = 2$
  and $\tau=\pi$. The Kerr state becomes a superposition of two
  coherent states: $|2e^{i\frac{\pi}{2}}\rangle$,
  $|2e^{i\frac{3\pi}{2}}\rangle$. }
\label{Wigner_3}
\end{center}
\end{figure}

Below we present the cat and kitten states - the superpositions of 2-6
coherent states - and their Wigner functions obtained from the Kerr
state (\ref{Kerr}) for some specific values of $\tau$ and $\alpha=2$,
Fig. \ref{Wigner_1}-\ref{Wigner_3}.

Setting $\tau = \frac{\pi}{3}$ the Kerr state becomes a superposition
of six coherent states
\begin{eqnarray}
|\Psi_K (\tau=\frac{\pi}{3})\rangle &=& c_1 |2e^{i\frac{\pi}{6}}\rangle + c_2
|2e^{i\frac{\pi}{2}}\rangle + c_3 |2e^{i\frac{5\pi}{6}}\rangle
\nonumber\\
&+& c_2 |2e^{i\frac{7\pi}{6}}\rangle
+ c_1|2e^{i\frac{3\pi}{2}}\rangle
+ c_4 |2e^{i\frac{11\pi}{6}}\rangle,
\label{cat_6}
\end{eqnarray}
with the following coefficients: $c_1 = \frac{1}{6}\left(2+2i +
e^{-i\frac{\pi}{6}} + e^{-i\frac{2\pi}{3}}\right)$, $c_2=
\frac{1}{6}\left(2-2i + e^{i\frac{5\pi}{6}} +
e^{-i\frac{2\pi}{3}}\right)$, $c_3 = \frac{1}{6}\left(1+i +
2e^{-i\frac{5\pi}{6}} + 2e^{i\frac{2\pi}{3}}\right)$, $c_4=
\frac{1}{6}\left(1 - i + 2e^{i\frac{\pi}{6}} +
2e^{i\frac{2\pi}{3}}\right)$.

 We have five coherent states for $\tau=\frac{2\pi}{5}$
\begin{eqnarray}
|\Psi_K(\tau=\frac{2\pi}{5})\rangle &=& c_1 |2\rangle + c_2 |2e^{i\frac{2\pi}{5}}\rangle + c_3
|2e^{i\frac{4\pi}{5}}\rangle + c_2 |2e^{i\frac{6\pi}{5}}\rangle + c_1 |2e^{i\frac{8\pi}{5}}\rangle,
\label{cat_5}
\end{eqnarray}
where $c_1 = \frac{1}{5}\left(2+2e^{i\frac{2\pi}{5}} +
e^{-i\frac{4\pi}{5}}\right)$, $c_2 =
\frac{1}{5}\left(2+2e^{-i\frac{2\pi}{5}} +
e^{i\frac{4\pi}{5}}\right)$, $c_3 =
\frac{1}{5}\left(1+2e^{i\frac{4\pi}{5}} +
2e^{-i\frac{4\pi}{5}}\right)$.

For $\tau = \frac{\pi}{2}$ we have superposition of four states
\begin{eqnarray}
|\Psi_K (\tau=\frac{\pi}{2})\rangle &=& c_1 |2e^{i\frac{\pi}{4}}\rangle + c_2
|2e^{i\frac{3\pi}{4}}\rangle + c_1 |2e^{i\frac{5\pi}{4}}\rangle - c_2 |2e^{i\frac{7\pi}{4}}\rangle,
\label{cat_4}
\end{eqnarray}
with $c_1 = \frac{1}{4}\left(2+e^{-i\frac{\pi}{4}} +
e^{i\frac{3\pi}{4}} \right)$, $c_2 = \frac{1}{2}e^{-i\frac{3\pi}{4}}$.

If $\tau = \frac{2\pi}{3}$ the Kerr state becomes superposition of
three states
\begin{eqnarray}
|\Psi_K(\tau=\frac{2\pi}{3})\rangle &=& c_1 |2\rangle + c_2 |2e^{i\frac{2\pi}{3}}\rangle + c_1
|2e^{i\frac{4\pi}{3}}\rangle, \label{cat_3}
\end{eqnarray}
where $c_1 = \frac{1}{3}\left(2+e^{i\frac{2\pi}{3}} \right)$, $c_2 =
\frac{1}{3}\left(1+2e^{-i\frac{2\pi}{3}} \right)$.

We achieve the usual cat state, the superposition of two coherent
states, for $\tau=\pi$
\begin{eqnarray}
|\Psi_K(\tau=\pi)\rangle &=& \frac{1}{2}(1-i) |2e^{i\frac{\pi}{2}}\rangle + \frac{1}{2}(1+i)
|2e^{i\frac{3\pi}{2}}\rangle. \label{cat_2}
\end{eqnarray}

\section{Approximated cat and kitten state generation using ion traps}

All the presented kitten states (\ref{cat_6}) - (\ref{cat_2}) can be
generated using current technology available for the ion traps and
already existing theoretical schemes for an ion arbitrary pure state
preparation, if the amplitude of the coherent state is not too large.

We show that the small kitten states can be very well approximated by
a superposition of only few Fock states with appropriate chosen
coefficients. It means that only a small number of Fock states in Eq.
(\ref{Kerr}) is of real significance and the sum can be cut off at
some $n=M$. We analyze the dependence of this number $M$ on the
initial coherent state amplitude $\alpha$.

A method of preparing an ion in a Paul trap~\cite{Leibfried2003} in a
finite superposition of Fock states with arbitrary coefficients
\begin{equation}
|\Psi^{(M)}\rangle_{\mathrm{ion}} = \sum_{n=0}^{M} c_n |n\rangle|g\rangle,
\end{equation}
has been proposed in \cite{Gardiner1997, Kneer1998}. In the above
formula $|n\rangle$ is an ion motional Fock state, defined according
to a harmonic oscillator potential in the trap, $M<\infty$. The states
$|g\rangle$ and $|e\rangle$ are the ion electronic ground and excited
states. The method is based on applying a series of laser pulses tuned
to the carrier frequency and the red sideband of the ion trap
alternately. It works both in and beyond the Lamb-Dicke regime.

Adjusting the time of laser pulses or their Rabi frequency we can obtain
\begin{equation}
c_n =\frac{1}{\sqrt{\sum_{k=0}^M \frac{|\alpha|^{2k}}{k!}}}
\frac{\alpha^n}{\sqrt{n!}}\, e^{i\frac{\tau}{2}n(n-1)}.
\label{coefficients}
\end{equation}
In that case we generate the approximated cat state applying $2M$ laser
pulses
\begin{equation}
|\Psi_K^{(M)}(\tau)\rangle = R_M C_{M-1} \cdot ... \cdot C_1 R_1 C_0
 |0,g\rangle.
\end{equation}

\begin{figure}[h!]
\begin{center}
\hbox to \textwidth{
\hss
\scalebox{0.7}{\includegraphics{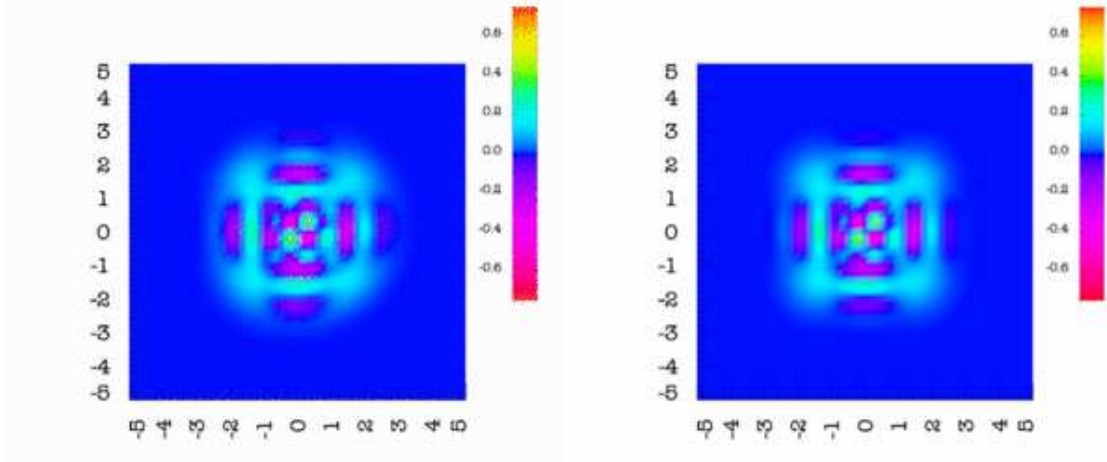}}
\scalebox{0.7}{\includegraphics{wd_a2_t0.5pi_M30d.eps}}
\hss}
\caption{(Color online) The Wigner function evaluated for $\alpha = 2$
  and $\tau=\frac{\pi}{2}$. The left figure: $M=10$. The
  right figure: $M=30$.}
\label{comparision}
\end{center}
\end{figure}

\begin{figure}[h!]
\begin{center}
\hbox to \textwidth{
\hss
\scalebox{0.7}{\includegraphics{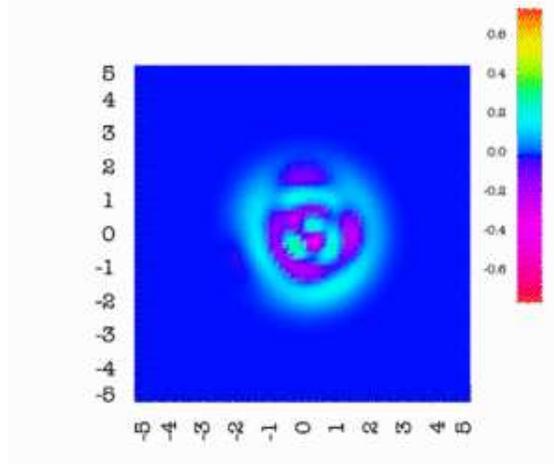}}
\hss}
\caption{(Color online) The Wigner function evaluated for $\alpha = 2$,
  $\tau=\frac{\pi}{2}$ and $M=5$.}
\label{comparision_2}
\end{center}
\end{figure}

The choice of the value $M$ of the cut off in the sum (\ref{Kerr}) is
based on comparison of the Wigner function computed for different
values of $M$.  On Fig. \ref{comparision} we present the Wigner
  function evaluated for $M=10$ and $M=30$ for $\alpha=2$ and $\tau=
  \frac{\pi}{2}$. Please note, that there is no significant difference
  between the left and right figure. We have also checked that $M=5$
  simplifies the Wigner function too much, Fig. \ref{comparision_2}.
Therefor, we assume that $|\Psi_K^{(10)}(\tau)\rangle$ approximates
the cat state for $\alpha=2$ good enough, which means that $20$ laser
pulses are required for its preparation.

The appropriate value of $M$ is approximately equal to the number of
significant number of coefficients in the sum (\ref{Kerr}). Below we
present the plots of absolute values of $c_n$,
Eq. (\ref{coefficients}), as a function of $n$ for $\alpha=1$,
$\alpha=2$ and $\alpha=5$. The number of the laser pulses required for
the kitten state preparation increases with the magnitude of the
amplitude $\alpha$ fast.

\begin{figure}[h!]
\begin{center}
\hbox to \textwidth{
\hss
\scalebox{0.55}{\includegraphics{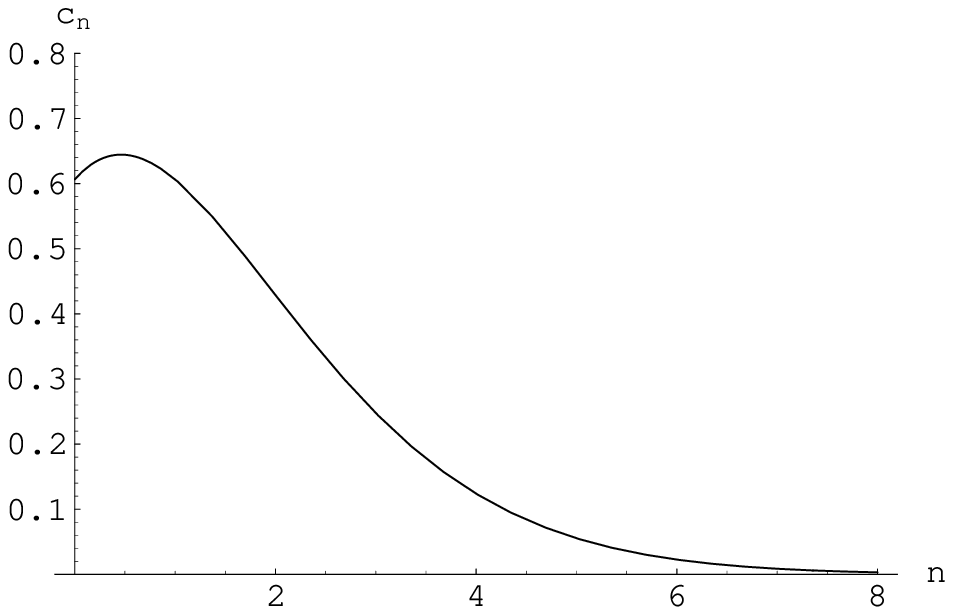}}
\scalebox{0.55}{\includegraphics{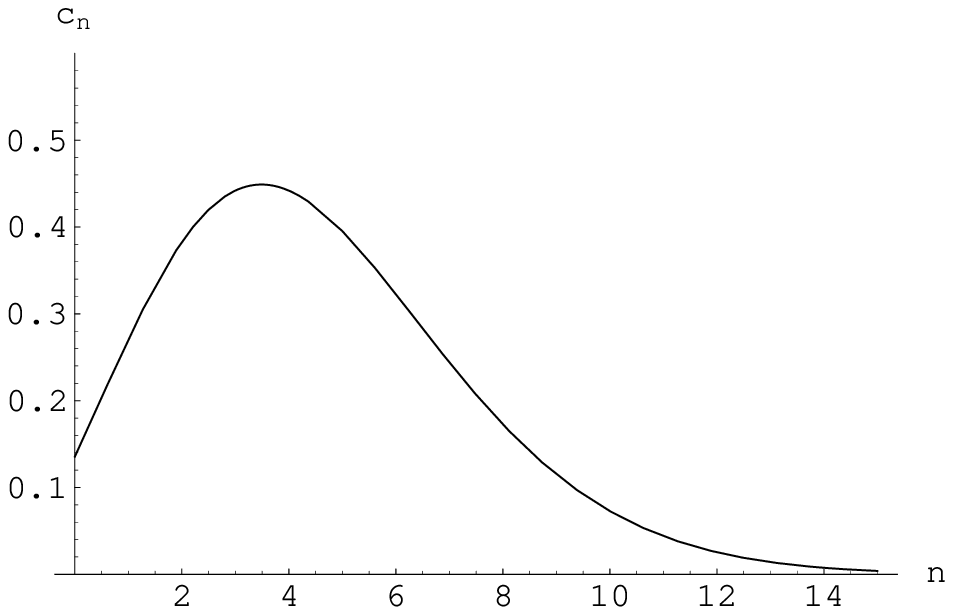}}
\scalebox{0.55}{\includegraphics{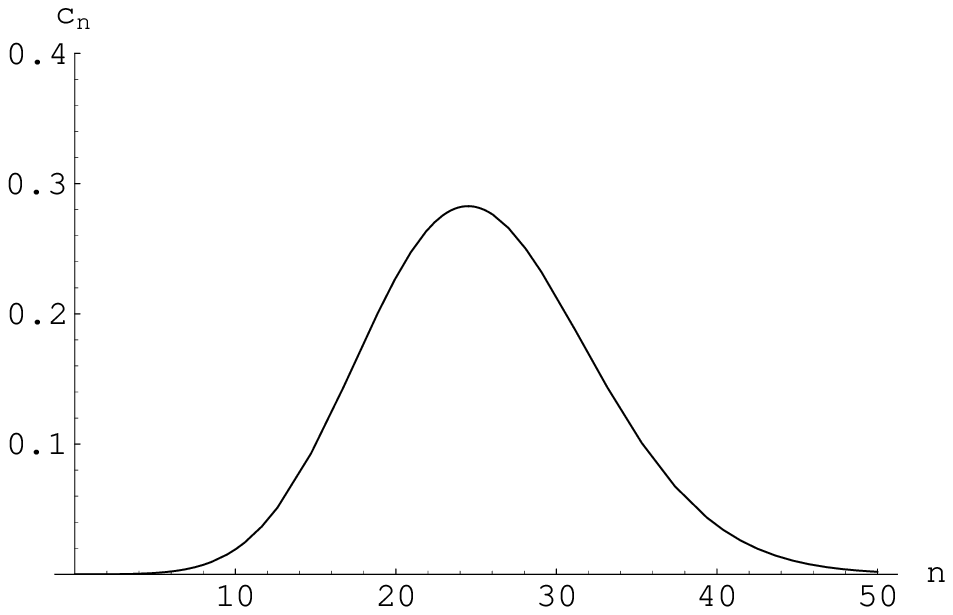}}
\hss}
\caption{The absolute values of $c_n$, Eq. (\ref{coefficients}), as a
  function of $n$ for $\alpha=1$ - the left figure, $\alpha=2$ - the
  middle figure and $\alpha=5$ - the right figure.}
\label{significant_coeff}
\end{center}
\end{figure}

The value of the amplitude eg. $\alpha=5$ requires about $100$ pulses
and the decoherence effects would have to be compensated for during
the preparation. Such an amplitude also requires higher number states
in Eq. (\ref{Kerr}) that have to be taken into account, which means
dealing with higher excitations of ion.

As an example, we list below the set of laser pulses parameters
required for approximated kitten state $|\Psi^{(10)}_K(\tau=
\frac{\pi}{2})\rangle_{\mathrm{ion}}$ for $\alpha = 2$ generation.

Assuming the carrier resonance Rabi frequency (for all pulses $C_i$)
equal to $\Omega_C = 1 \mathrm{MHz}$ and the red sideband Rabi
frequency (for all pulses $R_i$) equal to $\Omega_R = 100
\mathrm{kHz}$, the duration times and phases of pulses are as follows
\begin{displaymath}
\begin{array}{rllrll}
R10:& \phi = \pi,  & t_{10}^R = 995\mu\mathrm{s}; &
C9:& \phi = 0,     & t_9^C = 2.89\mu\mathrm{s},\\
R9:& \phi = 0.47,  & t_9^R = 387\mu\mathrm{s}; &
C8:& \phi = -0.83, & t_8^C = 1.16\mu\mathrm{s},\\
R8:& \phi = 7.23,  & t_8^R = 351\mu\mathrm{s}; &
C7:& \phi = 1.33,  & t_7^C = 1.30\mu\mathrm{s},\\
R7:& \phi = 4.26,  & t_7^R = 435\mu\mathrm{s}; &
C6:& \phi = 2.41,  & t_6^C = 2.21\mu\mathrm{s},\\
R6:& \phi = 5.00,  & t_6^R = 474\mu\mathrm{s}; &
C5:& \phi = -0.05, & t_5^C = 1.60\mu\mathrm{s},\\
R5:& \phi = 1.82,  & t_5^R = 546\mu\mathrm{s}; &
C4:& \phi = -0.86, & t_4^C = 2.44\mu\mathrm{s},\\
R4:& \phi = 2.21,  & t_4^R = 550\mu\mathrm{s}; &
C3:& \phi = -2.97, & t_3^C = 1.92\mu\mathrm{s},\\
R3:& \phi = -1.30, & t_3^R = 745\mu\mathrm{s}; &
C2:& \phi = -3.93, & t_2^C = 2.84\mu\mathrm{s},\\
R2:& \phi = -0.95, & t_2^R = 813\mu\mathrm{s}; &
C1:& \phi = -0.09, & t_1^C = 2.84\mu\mathrm{s},\\
R1:& \phi = 3.23,  & t_1^R = 1370\mu\mathrm{s}; &
C0:& \phi = -4.19, & t_0^C = 1.04\mu\mathrm{s}.
\end{array}
\end{displaymath}

We could also keep the duration time of pulses constant, $t_C = t_R =
1\mu\mathrm{s}$, and change the Rabi frequencies from pulse to pulse
\begin{displaymath}
\begin{array}{rllrll}
R10:& \phi = \pi, & \Omega_{10}^R = 94.9 \mathrm{MHz}; &
C9:& \phi = 0, & \Omega_9^C = 2.89 \mathrm{MHz},\\
R9:& \phi = 0.47, & \Omega_9^R = 36.7 \mathrm{MHz}; &
C8:& \phi = -0.83, & \Omega_8^C = 1.16 \mathrm{MHz},\\
R8:& \phi = 7.23, & \Omega_8^R = 33.1 \mathrm{MHz}; &
C7:& \phi = 1.33, & \Omega_7^C = 1.30 \mathrm{MHz},\\
R7:& \phi = 4.26,& \Omega_7^R = 40.7 \mathrm{MHz}; &
C6:& \phi = 2.41, & \Omega_6^C = 2.21 \mathrm{MHz},\\
R6:& \phi = 5.00, & \Omega_6^R = 43.9 \mathrm{MHz}; &
C5:& \phi = -0.05,& \Omega_5^C = 1.60 \mathrm{MHz},\\
R5:& \phi = 1.82, & \Omega_5^R = 49.9 \mathrm{MHz}; &
C4:& \phi = -0.86, & \Omega_4^C = 2.44 \mathrm{MHz},\\
R4:& \phi = 2.21, & \Omega_4^R = 49.1 \mathrm{MHz}; &
C3:& \phi = -2.97, & \Omega_3^C = 1.92 \mathrm{MHz},\\
R3:& \phi = -1.30, & \Omega_3^R = 64.5 \mathrm{MHz}; &
C2:& \phi = -3.93, & \Omega_2^C = 2.84 \mathrm{MHz},\\
R2:& \phi = -0.95, & \Omega_2^R = 66.4 \mathrm{MHz}; &
C1:& \phi = -0.09, & \Omega_1^C = 2.84 \mathrm{MHz},\\
R1:& \phi = 3.23, & \Omega_1^R = 96.9 \mathrm{MHz}; &
C0:& \phi = -4.19, & \Omega_0^C = 1.04 \mathrm{MHz}.
\end{array}
\end{displaymath}
The phases will not change.  As an initial state we take
$|\phi_{in}\rangle = (-0.97 + 0.25 i) |0,g\rangle$ and the Lamb-Dicke
parameter $\eta = 0.02$.

\section{Conclusion}

In this paper we have presented a method of an effective approximated
coherent superposition state generation for a single trapped ion in a
Paul trap.

At first, the ion is prepared in its ground state, both in motional
and electronic state. Then, step by step, the state is built up
applying a series of laser pulses tuned to the carrier resonance and
red sideband interaction alternately.

Fixing the amplitude of the coherent state, we can approximate the cat
state arbitrary well, increasing the number of applied pulses.

The cat and kitten states with their amplitude $\alpha \le 2$ are very
well approximated by a state which is available applying $20$ laser
pulses. The judgment is based on a Wigner function comparison.

\section*{Acknowledgments}

This work was partially supported by the  Grant PBZ-Min-008/P03/03.


\begin{thebibliography}{99}
\bibitem{Schroedinger} E. Schr\"odinger, Naturwissenschaften, {\bf
  23}, 807 (1935).
\bibitem{vanEnk} S. J. van Enk and O. Hirota, Phys. Rev. A, {\bf 64},
  022313 (2001).
\bibitem{Jeong} H. Jeong, M. S. Kim, and J. Lee, Phys. Rev. A, {\bf
  64}, 052308 (2001).
\bibitem{Jeong2} H. Jeong and M. S. Kim, Phys. Rev. A, {\bf 65},
  042305 (2002).
\bibitem{Ralph2003} T. C. Ralph, A. Gilchrist, G. J. Milburn, W. J.
  Munro and S. Glancy, Phys. Rev. A {\bf 68}, 042319 (2003).
\bibitem{Ralph2} T. C. Ralph, Phys. Rev. A, {\bf 65}, 042313 (2002).
\bibitem{Jeong3} H. Jeong et al., Phys. Rev. A, {\bf 67}, 012106
  (2003).
\bibitem{Buzek} V. Buzek et al., Phys. Rev. A, {\bf 45}, 6570 (1992).
\bibitem{Glancy} S. Glancy, H. M. Vasconcelos, and T. C. Ralph,
  Phys. Rev. A, {\bf 70}, 022317 (2004).
\bibitem{Monroe} C. Monroe, D. Meekhof, B. E. King, and
  D. J. Wineland, Science, {\bf 272}, 1131 (1996).
\bibitem{Myatt2000} Myatt et. al, Nature {\bf 403}, 269 (2000).
\bibitem{Brune} M. Brune et al., Phys. Rev. Lett., {\bf 77}, 4887
  (1996).
\bibitem{Jeong4} H. Jeong, M. S. Kim, T. C. Ralph, and B. S. Ham,
  Phys. Rev. A, {\bf 70}, 061801(R) (2004).
\bibitem{Stobinska} M. Stobi\'nska, G. J. Milburn, K. W\'odkiewicz,
  quant-ph/0605166.
\bibitem{Blatt2005} R. Blatt and A. Steane, {\em Quantum
  Information Processing and Communication in Europe\/}, pp. 161-169,
  European Communities (2005).
  \bibitem{Ekert1996} A. Ekert and Josza, Rev. Mod. Phys. {\bf 68},
    733 (1996).
  \bibitem{Roos1999} Ch. Roos et al., Phys. Rev. Lett. {\bf 83}, 4713
    (1999).
  \bibitem{Meekhof1996} D. M. Meekhof et al., Phys. Rev. Lett. {\bf
    76}, 1796 (1996).
\bibitem{Heinzen1990} D. J. Heinzen and D. J. Wineland, Phys. Rev. A
  {\bf 42}, 2977 (1990).
\bibitem{Tanas} R. Tana\'s, {\em Nonclassical states of light
  propagating in Kerr media\/}, in Theory of Non-Classical States of
  Light, V. Dodonov and V. I. Man'ko eds., Taylor and Francis, London
  2003.
\bibitem{Bialynicka} Z. Bialynicka-Birula, Phys. Rev. {\bf 173}, 1207
  (1968).
\bibitem{Averbukh1989} I. Sh. Averbukh and N. F. Perelman, Phys. Lett.
  {\bf 139}, 449 (1989).
\bibitem{Leibfried2003} D. Leibfried, R. Blatt, C. Monroe, and D.
  Wineland, Rev. Mod. Phys. {\bf 75}, 281 (2003).
\bibitem{Gardiner1997} S. A. Gardiner, J. I. Cirac, and P. Zoller,
  Phys. Rev. A {\bf 55}, 1683 (1997).
\bibitem{Kneer1998} B. Kneer and C. K. Law, Phys. Rev. A {\bf 57},
  2096 (1998).
\end{thebibliography}
\end{document}